\documentclass[12pt,letterpaper]{article}
\usepackage{amsmath,amssymb,pgf,pgfarrows,pgfnodes,float,appendix, hyperref}
\usepackage{graphicx}
\usepackage{subfigure}
\usepackage[margin=0.9in]{geometry}

\def\Q{{\cal Q}}

\title{{\rm\footnotesize \qquad \qquad \qquad \qquad \qquad \ \qquad \qquad \qquad \ \ \ \ \ \                  RUNHETC-2018-25,  UTTG 15-19}\vskip.5in Cosmological Implications of the Bekenstein Bound}
\author{Tom Banks\\
Department of Physics and NHETC\\
Rutgers University, Piscataway, NJ 08854\\
E-mail: \href{mailto:banks@physics.rutgers.edu}{banks@physics.rutgers.edu}
\\
\\
Willy Fischler\\
Department of Physics and Texas Cosmology Center\\
University of Texas, Austin, TX 78712\\
E-mail: \href{mailto:fischler@physics.utexas.edu}{fischler@physics.utexas.edu}}

\date{September 25, 2018}
\begin{document}
\maketitle

\begin{abstract}
A brief review of ``Holographic Space-Time'' in the light of the seminal contributions of Jacob Bekenstein. \end{abstract}

\section{Introduction}

Jacob Bekenstein's revolutionary discovery of the entropy of black hole horizons has important implications for cosmology and, in the opinion of the present authors, may provide the key clue to a general quantum theory of gravity. The present article will review attempts to extend Bekenstein's ideas to cosmological, and more general space-times.  The logical line of argument runs from Bekenstein and Hawking\cite{Bekhawk}, through the Gibbons-Hawking discovery of de Sitter temperature\cite{Gibbhawk}, the fundamental results of Jacobson\cite{ted}, and the initial cosmological conjecture of Susskind and one of the present authors\cite{fsb}, which led to the general conjecture of Bousso\cite{raphael}.  We view our own work on Holographic Space-time (HST) to be the natural extension of these ideas\cite{hst}.  There have been other attempts to generalize Bekenstein's results to cosmological space-times\cite{bruvenetal}, but they appear to lie outside the general pattern that emerges from this historical development.

Bekenstein's assignment of entropy to black holes implied a non-zero temperature and Hawking's revolutionary calculation of the thermal spectrum of black holes was verification of Bekenstein's conjecture.  This is sometimes taken to imply that quantum field theory must be a good description near the horizon of a large black hole, but as shown by the work of \cite{hhgh} the key to the thermal nature of black holes is the periodic identification of the Euclidean time coordinate in the non-singular continuation of black hole metrics to Euclidean signature.  This implies thermal behavior of any quantum system using that time coordinate.  The latter argument is the simplest way to understand the Gibbons-Hawking temperature of de Sitter space\footnote{Since there is no general argument that different dS spaces can be thought of as states of a single system, it's unclear if one should write the thermodynamic relation $dE = T dS$ in this context, but if one does, and uses the Bekenstein relation between horizon area and entropy, one deduces the correct energy density of dS space.}.

The first clue about extending Bekenstein's conjecture to general space-times came more than two decades later, in a paper by T. Jacobson\cite{ted}. In our opinion, the seminal work of Bekenstein, and its elaboration by Hawking, Gibbons and Jacobson, will eventually be seen as laying the foundations of a general quantum theory of gravity.  Jacobson considered a spacetime whose radii of curvature are larger than all microscopic scales, and argued that there was consequently a large region of spacetime which was approximately flat.
He then makes three postulates (one only implicit)

\begin{itemize}

\item One can localize the energy and entropy density in small regions and use hydrodynamic (localized thermodynamic) equations.

\item Bekenstein's law relating area to entropy is valid for infinitesimal changes in area, in particular those which are described by Raychauduri's equation and related to the Einstein curvature.

\item Unruh's relation between acceleration and temperature in flat space is valid in the limit of infinite acceleration.  There is a lot hidden in this assumption.  It implies in particular that the entropy that appears in thermodynamic relations is the maximal entropy allowed for the system. In the context of quantum mechanics this means that the entropy should be interpreted as the logarithm of the dimension of a localized quantum system.  

\end{itemize}

The thermodynamic relation
\begin{equation} dE = T dS , \end{equation}
can be written in terms of the change in area of a causal diamond as a consequence (for example) of the change in proper time between its tips.
\begin{equation} 4 dE L_P^2 = \int_{\bf H} \theta d\lambda d{\bf A} . \end{equation}
We are thinking about an Unruh trajectory that is near the boundary of the diamond, with its inflection point nearest to a point $p$ on the boundary. $\chi^a$ is the approximate boost Killing vector (here we assume low curvature) that generates this trajectory :  $\chi^a = - \kappa \lambda k^a $, where $\kappa$ is the acceleration of the trajectory where the norm of $\chi^a$ is $1$, $k^a$ is the tangent to the generators of the local Rindler horizon, and $\lambda$ is an affine parameter that vanishes at $p$ and is negative to the past of $p$.   $d{\bf A}$ is the change in area on the horizon.  
We take $\kappa$ to be very large, so that the trajectory stays near the point $p$.
That means that we are taking the infinite temperature limit since $T = \kappa / 2\pi$.  In this limit $\kappa$ appears linearly on both sides of the first law of thermodynamics since
we also have
\begin{equation} dE = - \kappa \int_{\bf H} \lambda T_{ab} k^a k^b d\lambda d{\bf A} .\end{equation}
Note that this is proportional to the time dependent Hamiltonian for motion along the trajectory.
Since $p$ is on the maximal area surface at the boundary of the diamond the expansion and shear vanish at $p$ and the Raychauduri equation is approximately \begin{equation} d\theta = - R_{ab}k^a k^b d\lambda , \end{equation} which can be integrated to find $\theta$ near $p$.  Thus we get
\begin{equation} 8\pi L_P^2 \int_{\bf H} T_{ab} k^a k^b d\lambda d{\bf A} =   \int_{\bf H} R_{ab} k^a k^b d\lambda d{\bf A} . \end{equation}   If we impose covariant conservation of the stress tensor, then this can only be true if Einstein's field equation holds for every point in space time, when dotted twice into every local null vector at that point. Consequently, one gets Einstein's equation with an undetermined cosmological constant.  

Our interpretation of the latter fact is that the cosmological constant is {\it not } a local energy density.  We conjectured long ago that the c.c. was instead an asymptotic boundary condition.  A causal diamond has two large scale geometric properties: the proper time of the geodesic between its tips, and the area of its holographic screen.  The c.c. controls the relation between the two limits where these quantities become large.  For positive $\Lambda$ the area remains finite for proper time going to infinity. For negative $\Lambda$ the area goes to infinity at finite proper time.  For vanishing $\Lambda$ we have $A \sim t^{d-2}$ as $t\rightarrow\infty$.  

Jacobson's result shows that Einstein's field equations are the hydrodynamic equations of the BHJFSB entropy law for causal diamonds. String theory tells us that we can think of all fields in physics as originating from the supersymmetric partners of and other features of the metric in high enough dimension.  Thus, all of field theory is hydrodynamics.  This is a clue to the nature of quantum gravity.  In familiar physics, one quantizes the equations of hydrodynamics only near the ground state of a system where all low energy excitations are quantized hydrodynamic fields (including the fermions of Fermi liquid theory).  These are linear field theories, with small perturbative corrections. The classical non-linear equations of hydrodynamics are valid instead in high entropy situations where the microscopic quantum states have nothing to do with phonons.  Entropy appears in these equations, but is not explained by them.   Similarly we expect that the entropy of horizons will come from quantum systems very different from bulk quantum field theory.  

We also reiterate that Jacobson's use of infinite Unruh temperature systems to derive his result, implies that the entropy law applies to the full dimension of the quantum Hilbert space, not some less random density matrix.

\section{The Fischler-Susskind Bound}

In the early spring of 1998, Fischler and Susskind were wondering whether the spectacular discovery of Jacob Bekenstein in the context of black holes could hold for general space times. Their prejudice at the time was that indeed this new paradigm should apply more generally.  In particular, cosmological space-times seemed a fertile and especially useful arena in the search for a first step in generalizing Bekenstein's idea.  This led them to consider flat FRW universes as simple geometries where one might be able to extend the Bekenstein entropy bound.

At that time we had learned from Matrix Theory and what later became known as the AdS-CFT connection, that the holographic principle, requires that the degrees of
freedom of a spatial region reside not in the interior as in an ordinary quantum field theory but on the surface of the region. Furthermore it requires the number of
degrees of freedom per unit area to be no greater than 1 per Planck area. As a consequence, the entropy of a region must not exceed its area in Planck units. The task at hand was how to generalize this principle to cosmology. What follows is a brief summary of the Fischler-Susskind paper on the holographic principle in cosmology \cite{fsb}

The FRW metric has the usual form 
\begin{equation} ds^2 = dt^2 - a^2(t) dx^idx^i \end{equation}

The number of spatial dimensions will be kept general so that $ i$ runs from $1$ to $d$.
The cosmological assumptions are the usual ones including homogeneity, isotropy and in the late universe constant entropy density in comoving coordinates.

A first attempt was very quickly ruled out: the entropy in any region of coordinate size $ \Delta x \sim R$ never exceeds the area. Since our
assumptions require the entropy density to be constant, the entropy in a region is proportional to its coordinate volume $R^d$. Furthermore in flat space the surface area of the region  
grows with $R$ like $[Ra(t)]^{d-1}$. Obviously when $R$ becomes large enough, the entropy exceeds the area and the principle is violated.

A more sophisticated version goes as follows. Consider a spherical spatial region $\Gamma$ of coordinate size $R$ with boundary $B$. Now consider the light-like  
surface $L$ formed by past light rays from $B$ toward the center of $\Gamma$. One can consider then three situations. If $R$ is smaller
than the coordinate distance to the cosmological horizon $R_H$ then the surface $L$ forms a light cone with its tip in the future of the singularity at $t=0$. If $R$
is the size of the horizon then $L$ is still a cone but with its tip at $t=0$. However if $R>R_H$ the surface is
a truncated cone. Now consider all the entropy (particles) which pass through $L$. For the first two cases this is the same as the entropy in the interior of  
$\Gamma$ at the instant $t$.
But in the last case the entropy within $\Gamma$ exceeds the entropy passing through $L$. The proposed holographic principle is that the entropy passing through $L$ never exceeds the area of the bounding surface $B$. It is not difficult to see that for the homogeneous case, this reduces to a single condition:

The entropy contained within a volume of coordinate size $R_H$ should not exceed the area of the horizon in Planck units. In terms of the (constant) comoving  
entropy density $\sigma$ 

\begin{equation} \sigma R_H^d <[ aR_H]^{d-1}\end{equation}

with every thing measured in Planck units. Note that both $R_H$ and  $a$ are functions of time and that this equation must be true at all time.

Let us first determine whether the bound is true today. The entropy of the observable universe is of order $10^{86}$ and the horizon size (age) of the universe is of order
$10^{60}$. Therefore the ratio of entropy to area is much smaller than 1. Now consider whether it will continue to be true in the future. Assume that $a(t) \sim t^p$. The horizon size is
determined by
\begin{equation} R_H(t) = \int_0^t {dt \over a(t)} \sim t^{1-p} \end{equation}
Thus in order for this equation to continue to be true into the remote future we must satisfy
\begin{equation} p>{1 \over d}\end{equation}

In other words there is a lower bound on the expansion rate.

The bound on the expansion rate is easily translated to a bound on the equation of state. Assume
that the equation of state has the usual form
\begin{equation} P = \gamma \epsilon \end{equation}
where $P$ and $\epsilon$ are pressure and energy density. Standard methods yield a solution of the Einstein equations
\begin{equation} a(t) \sim t^{2 \over d(1+ \gamma)} \end{equation}
Thus $p ={2 \over d(1+ \gamma)}$ and the inequality for the equation of state becomes 
\begin{equation} \gamma < 1 \end{equation}

This bound is well known and follows from entirely different considerations. It describes the most incompressible fluid that is consistent with special relativity. A violation
of the bound would mean that the velocity of sound exceeds the velocity of light. We will take this agreement as evidence that our formulation of the cosmological holographic principle
is on the right track.

Although violating the bound  on the equation of state is impossible, saturating it is easy.  We will give two examples.  In the first example the energy density of the universe is dominated by a homogeneous massless minimally coupled scalar field. It is well known that in this case the pressure and energy density are equal \footnote{In HST cosmology this equation of state describes the early stage of the universe \cite{holcosnew}}. 

Another example involves flat but anisotropic universes, for example the anisotropic \cite{kas} universes with metric:
\begin{equation} ds^2 = dt^2 - \Sigma_i\ {t^{2p_i} {dx_i}^2} \end{equation}\

The ratio entropy/area, $S/A$ ,for this case is easily evaluated to be:

\begin{equation} S/A = \Pi_i \ R_{H,i}~ / {[\Pi_j\ {t^{p_j}t^{1-p_j}]}^{{d-1}/d}} \end{equation}

where $R_{H,i} = t^{1-p_i}$ is the coordinate size of the horizon in
direction i . The denominator on the left hand side of this equation is the proper area of the horizon.

It follows that:

\begin{equation} S/A = t~ ^ {1 - \Sigma_i\ p_i } \end{equation}

 The conditions on the exponents of the Kasner solutions are obtained by using the Einstein equations.The
exponents  satisfy the following equations:

\begin{equation} \Sigma_i\ p_i = 1 \end{equation}

\begin{equation} \Sigma_i\ p_i^2 = 1 \end{equation}

It follows from these equations that $ S/A $ for these flat anisotropic universes is constant in time.Thus depending on the boundary condition the holographic bound may be saturated by these universes.  Note that in these "models" we do not really explain the entropy of the system.  The scalar field/Kasner metric acts like a hydrodynamic variable, which encodes the entropy but does not explain it\footnote{Indeed, we get the same equation of state as these models in a kinetic dominated, or mixmaster universe, which has gravitational/scalar fields that depend on position but have no spatial gradient terms in their action.  These models have {\it too much} entropy to be compatible with the cosmological version of Bekenstein's bound.}.

Having established that the holographic principle will not be violated in the future we turn to the past. First consider the entropy-area ratio ($\rho = S/A$) at the time of decoupling.
Standard estimates give a ratio which is about $10^6$ times bigger than todays ratio. 
Thus at decoupling 
\begin{equation} \rho(t_d) \sim 10^{-28} \end{equation}
During a radiation dominated era it is easily seen that $\rho$ is proportional to $t^{- 1/2}$.

\begin{equation} \rho = 10^{-28} \left[{t_d \over t}\right]^{1/2} \end{equation}
Remarkably, in Planck units $t_d^{1/2} \sim 10^{28}$ so that $\rho < 1$ for all times later than the Planck time. The entropy in the universe is as large as it  
can be without the holographic principle having been violated in the early universe!

When Fischler and Susskind tried to extend their approach to closed FRW universes they found that their arguments led to a violation of the bound for such cosmologies. This puzzle was resolved, and a much more general formulation of a Bekenstein like bound discovered, in the work of Bousso\cite{raphael}

\section{The Bousso Bound}

The work by Fischler and Susskind inspired Bousso to generalize the bound and express it in a covariant language.  The result was Bousso's magnificent general conjecture, which he called the Covariant Entropy Bound.  A detailed description of this can be found in Bousso's contributions to this volume.  We will only mention a few differences of emphasis, which have developed in our extensive use of Bousso's work in our own.  

Bousso's bound applies to a general space-like two surface and is described in terms of conditions on the light sheets emanating from it.  A particular case of such a two surface is associated with a causal diamond.  The boundary of the diamond is the intersection of two light cones, which form a continuous, but not differentiable manifold.  On each light cone there is a metric of the form
\begin{equation} ds^2 = (du - A_i ({\bf x} , u) dx^i) B_j ({\bf x} , u) dx^j + g_{ij} ({\bf x}, u) dx^i dx^j . \end{equation} The area of the space-like two surfaces at fixed $u$ increases as one moves away from the tips of the diamond.  By continuity, there is a maximum value of the area for some value of $u$, which might be at the joining point of the two diamonds.  The area of that maximal area surface satisfies the Covariant Entropy Bound of Bousso.  Bousso's conjecture says that it bounds four times the entropy of states localized in the causal diamond.  In our work, we've used the phrase "holographic screen of the diamond", to refer to this maximal area surface.  In Bousso's work the name {\it holographic screen} refers to the full null history of this $2$-surface.  

There is a well known theorem in Lorentzian geometry, which says that a metric is completely determined in terms of properties of its collection of causal diamonds.  For any pair of diamonds, there is a maximal area diamond in it's intersection, and the causal structure of the manifold is completely determined by specifying the collection of diamonds and their intersections. The conformal factor is determined in terms of the areas of all the diamonds. If, in addition, we use the Bousso-Jacobson generalization of Bekenstein's area law to arbitrary causal diamonds, and assume that the entropy bound is in fact saturated for the maximally uncertain density matrix
associated with a causal diamond, then we can translate a Lorentzian geometry into a statement about quantum Hilbert spaces.  

Indeed, we have long known how to translate causal relations into properties of a quantum operator algebra\cite{AQFT}.  Bekenstein's seminal work, as generalized by Jacobson, Fischler-Susskind, and Bousso, completes the process of translation.  The more refined version of Bousso's conjecture, which insists that the entropy bound is saturated, is best justified by Jacobson's use of infinite temperature Unruh trajectories to derive Einstein's equations from hydrodynamics.  

It's an immediate consequence of these considerations that the ancient idea of quantum gravity as a theory of fluctuating geometries is not correct.  Jacobson's arguments view the geometry of space-time as a hydrodynamic variable.  The more refined quantum version of this idea that the present authors have proposed in the theory of Holographic Space-time (HST) views geometry as an emergent description relating the dimensions of Hilbert spaces nesting inside each other.  There is no quantum interference involved.

One can understand how these ideas are consistent with string theory's match onto low energy effective field theory by recalling that in ordinary condensed matter physics, hydrodynamic equations are useful in two quite different regimes.  For systems with a ground state, hydrodynamic variables associated with conserved symmetry currents often give a complete quantum mechanical description of the low energy excitations of the system.  They give rise to a set of gapless boson and fermion fields, which are weakly interacting.  The dominant longwavelength interactions are described by generalized Navier Stokes equations because those equations follow from symmetries and the derivative expansion.   Recently, two groups have presented general effective field theory descriptions of the hydrodynamic modes of relativistic quantum field theories\cite{liurangamani}.  

The traditional use of hydrodynamics however is in a completely different regime, high in the energy spectrum, where the density of states is very large.  In this regime, hydrodynamic variables satisfy non-linear classical equations, and the quantum fluctuations around those solutions are exponentially small as a function of the entropy density.  The microscopic quantum fluctuations of the system take place on a wide variety of time scales, between the natural microscopic scales $\tau$ and $e^{S} \tau$.  They have nothing to do with quantum fluctuations of the hydrodynamic fields, but rather with short wavelength fluctuations, which do not change the hydrodynamic variables by perceptible amounts, because the latter are coarse grained volume averages.

Jacobson's derivation of the Einstein's equations takes place in this high entropy regime.  The novel feature of theories of gravity are connected to the fact that when the c.c. is non-negative, this high entropy regime is one of extremely low energy.  For a causal diamond of radius $R$, for positive c.c. or $ R \ll R_{AdS}$ for negative c.c., typical states contribute to the Hamiltonian an amount of order $R^{-1}$.   In the limit $R/L_P \rightarrow\infty$, in flat space, these states become "soft gravitons".  They are topological degrees of freedom, which decouple from the scattering operator of finite energy states.

The traditional methods of string theory hide these states in two different ways.  In perturbative string theory in Minkowski space, amplitudes for producing large numbers of soft gravitons in processes initiated by a few finite energy particles, are very small at low energy\footnote{Here we are implicitly assuming a Minkowski space of sufficiently high dimension.  It's well known that in four dimensions the gravitational scattering operator has vanishing matrix elements between Fock space states even in perturbation theory. We believe that in all dimensions Fock space is a subspace of  a representation space of an asymptotic current algebra, generated by the action of currents that are delta functions in angle, and that the scattering operator is not unitary on this subspace\cite{tbscatt}.} .  At energies of order string scale, long strings, the precursors\cite{horpol} of black holes, are produced even in few particle collisions.  These large meta-stable objects emit infinite numbers of soft gravitons in their evolution and decay.   The non-perturbative approach to string theory in AdS spaces, is a system with a gap, and the asymptotic soft graviton states are simply banished from the Hilbert space. High entropy is encountered, as in non-gravitational physics, only far above the gap.  As a consequence\cite{tbwfads} gravity in AdS space is a poor guide to local physics and cosmology.

Thus, traditional string theory deals only with the low entropy part of the spectrum of a quantum gravitational Hamiltonian, where quantized hydrodynamics is a good approximation to much of the physics.   Classical gravity, like the Navier-Stokes equation, is still a good description of coarse grained properties of high entropy states, but it is a mistake to quantize it in those situations.  The theory of Holographic Space Time (HST) takes the JFSB generalization of Bekenstein's law as a fundamental principle and gives a transparent model of the physics of horizons and inflation.  We turn next to a brief description of the principles of HST and its application to inflationary cosmology.

\section{The HST Formalism}

Consider a causal diamond $D_1$ in a general space-time, embedded in a larger causal diamond, $D_2$ whose holographic screen has finite area.  The BJFSB principle implies that each of these diamonds is associated in the quantum theory with a finite dimensional operator algebra.  The notion of causality implies that the smaller algebra ${\cal A}_1$ has a nontrivial commutant inside the larger one.  Otherwise we would have a genuine EPR paradox, with quantum information being transmitted faster than the speed of light.  For finite dimensions, this implies that \begin{equation} {\cal A_1} \otimes \bar{{\cal A}}_1 , \end{equation} and that the Hilbert space has a similar factorization.
This generalizes to diamonds that are not nested, but share an intersection region (\ref{fig. 1}). That region contains a maximal area causal diamond, which is represented in quantum mechanics as a tensor factor in the individual diamond Hilbert spaces.  Consistency of the two diamonds' descriptions of the intersection implies that the state on individual diamond Hilbert spaces always gives a density matrix for the intersection diamond $D_{12}$ with the same eigenvalue spectrum.  This is the quantum analog of Einstein's general principle of relativity, as we will see in a moment.  The unitary transformations relating the two descriptions of the density matrix are quantum analogs of general coordinate transformations between different physical gauges.  We therefore call these consistency requirements {\it Quantum General Coordinate Invariance} (QGCI).

\begin{figure}[h!]
\begin{center}
\includegraphics[width=12cm]{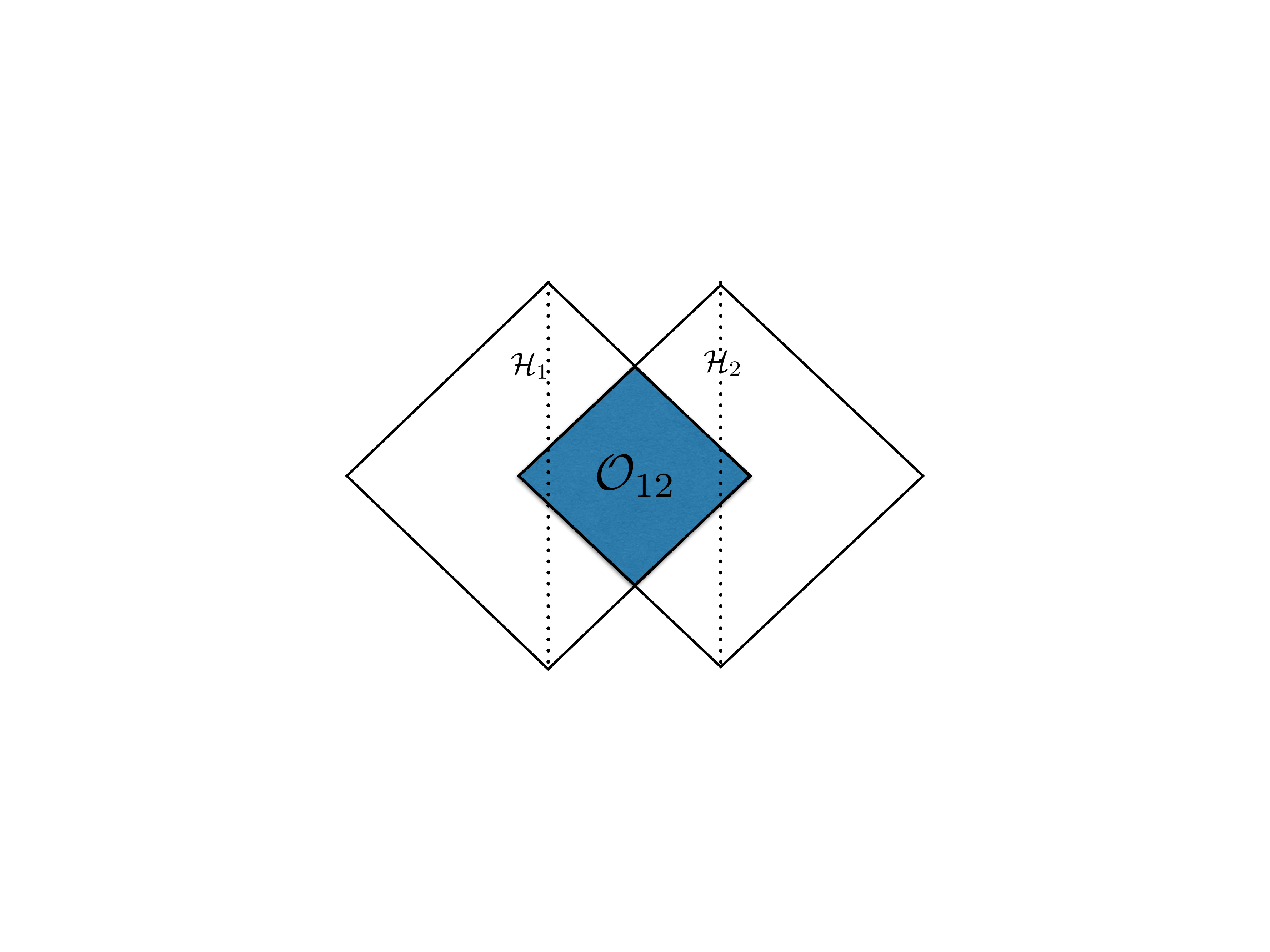}
\end{center}

\caption{ 
Intersecting diamonds}

\label{fig1}
\end{figure}

A causal diamond in space-time is characterized by two intrinsic geometrical parameters, the maximal proper time between its tips, and the maximal area on its boundary.  In HST, the latter is encoded in the dimension of a Hilbert space, while the former is directly connected with the time evolution in the quantum system.   We will have a different quantum description for every time-like trajectory (though we'll stick to geodesics in this article).  Choosing a set of nested proper time intervals along a trajectory is equivalent to choosing a nested set of causal diamonds, and part of the geometrical information is encoded in the relation between the proper time dependence of the Hamiltonian and the growth of the area.  In order to be consistent with QGCI the time evolution over any interval $I$, must factorize into $U_{in} U_{out}$ where $U_{in}$ is constructed from operators in ${\cal A}(I)$ and $U_{out}$ from operators in $\bar{{\cal A}} (I)$.  Thus, the Hamiltonian {\it must} be time dependent.  This is a reflection of the geometrical fact that any time slicing that remains within a causal diamond has a generator which is not a Killing vector of space-time, even in space-times with time-like Killing vectors.  Figure 2 shows the relation between the causal coordinates of HST and the standard FRW coordinate systems of cosmology.  Note that in any such coordinate system there is a large redshift between the proper time along a geodesic and the local times of degrees of freedom that are localized on the boundary.  This is incorporated in the Hamiltonian in a relative factor of $1/N$ between terms describing the boundary and terms describing excitations localized in the bulk.  $N $ scales like $A^{1/2}/ L_P$ when that quantity is large.  Note finally that associating the Hamiltonian with proper time along particular trajectories completely resolves the "Problem of Time" in GR.  Relativity is imposed as consistency conditions between quantum systems with different Hamiltonians, rather than insisting that any particular description be invariant.

\begin{figure}[h!]
\begin{center}
\includegraphics[width=12cm]{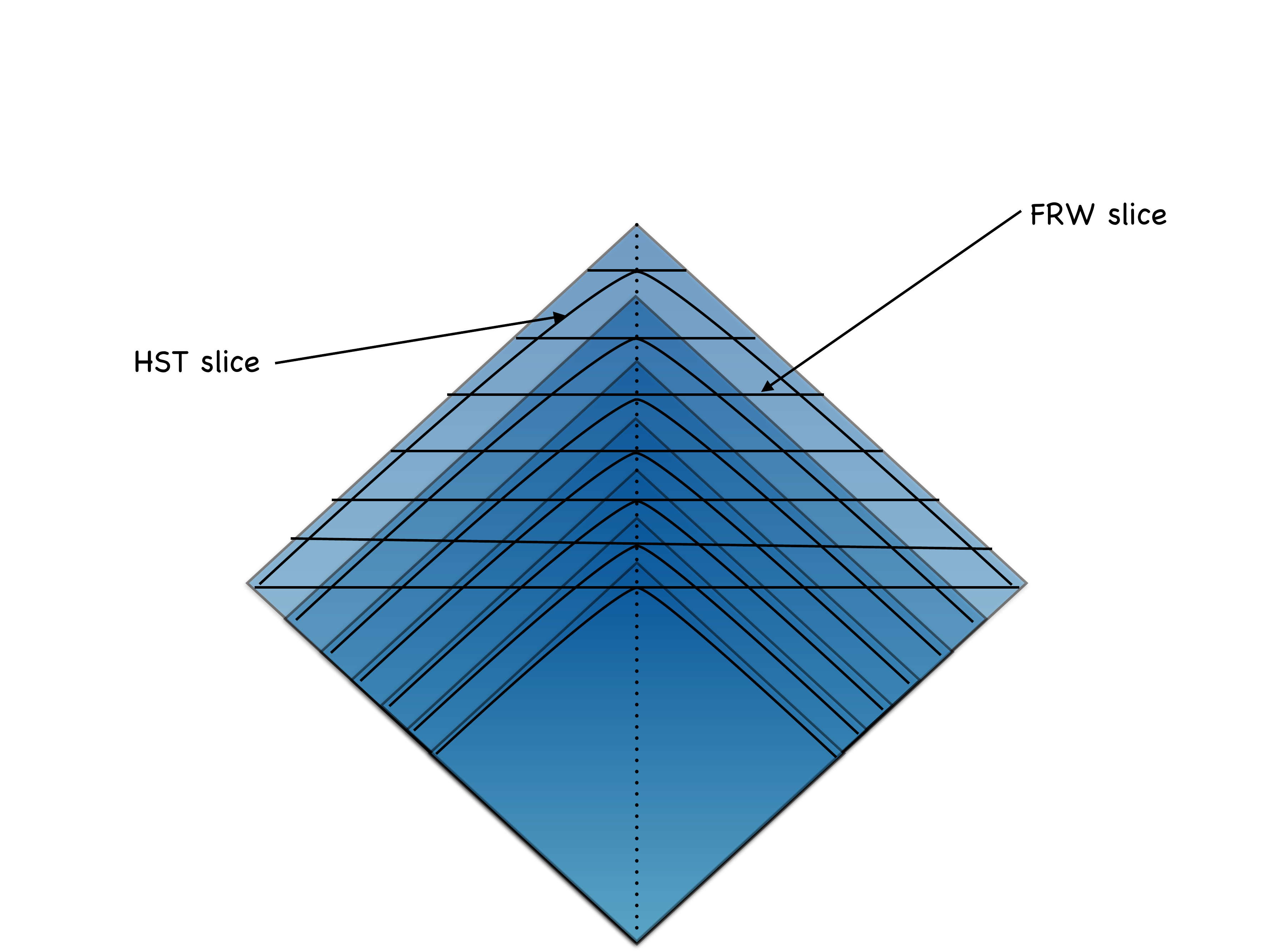}
\end{center}

\caption{ 
HST and FRW coordinates}

\label{fig2}
\end{figure}

Although this is not the historical line of argument, one can motivate the explicit models we study, by two observations about black holes.  The first is that inserting a black hole into de Sitter (dS) space {\it reduces} the area of the horizon and thus the entropy of the system.  In fact, the Schwarzschild dS metric gives a completely classical (except for its use of Bekenstein's quantum identification of entropy and area) derivation of the Gibbons-Hawking temperature of this space-time.  The value of that temperature suggests that the states responsible for the entropy of empty dS space contribute to the Hamiltonian of a static geodesic via an operator $1/N H_B$, where $H_B$ has an $N$ independent bound.  Objects localized near the geodesic will have energies that are $o(N^0)$.  One is also led to describe localized states by a number of constraints of order $EN$ where each constraint operator acts on a Hilbert space whose size is independent of $N$.  Indeed a randomly chosen state will satisfy this many constraints with quantum probability $e^{-c EN}$.  $E$ should thus be thought of as proportional to the energy of the state in Planck units and must be $\ll N$ if the state is to be meta-stable. 

The second important property of black holes that we will use is that they are fast scramblers\cite{hpss}. Information localized on the horizon is scrambled in a time of order the logarithm of the entropy in horizon radius units.  This is a property of all black holes in Minkowski or dS space, but in AdS space, the scrambling is only ballistic (a power of the entropy) when the black hole radius exceeds the AdS radius.   This follows both from the nature of the quasi-normal modes\cite{hh} and from the fact that the quantum theory is a local field theory\footnote{More precisely, the scrambling is fast out to the AdS radius, and then becomes ballistic at larger distances.  Note that the infall time to the singularity is always the AdS radius and does not increase with the Schwarzschild radius as it does for non-negative c.c. .}.

To motivate the choice of variables and Hamiltonians in HST models, we think about scattering theory in asymptotically flat space (AfS). It's convenient to describe the conformal boundary of AfS by two copies of the momentum light cone $P^2 = 0$, one at each angle for massless momenta penetrating the boundary and one for momenta flowing along the boundary.  We need the latter to describe currents carried by massive particles.  We can describe everything at infinity by currents carrying quantum numbers through the sphere at infinity.  These are generalized functions $J_i (P)$. Our own world contains particles of half integer spin, so among the these there must be spinors $\Q_{\alpha}^i (P)$, which are null plane spinors for $P$ and perhaps also null plane spinors for the reflected null vector $\tilde{P}$.  The spin statistics connection implies that the $Q$ operators at different points must anti-commute.  In fact, at coincident points they satisfy a generalization of the Awada-Gibbons-Shaw algebra\cite{ags}, but we will not have to use that.  The $Q_{\alpha}^i$ carry quantum numbers sufficient to describe everything in space-time so it's reasonable to assume that they generate the entire operator algebra.

Now imagine a finite causal diamond associated with a finite proper time interval along a particular geodesic.  We have to find a finite dimensional subalgebra of variables compatible with rotation invariance around that geodesic, and the only way to do that is to impose an angular momentum cutoff.  That is, we take the operator algebra of the diamond to be generated by $q_{ab}^i = - q_{ba}^i$, where the indices run from $1$ to something of order $N = R/L_P$\footnote{This gives us all half integer representations of $SO(3)$ up to the cutoff.}  Rotation invariance of the commutation relations implies that $[q_{ab}^i , q_{cd}^j ]_+ = (\delta_{ac}\delta_{bd} - \delta_{ad}\delta_{bc}) B^{ij} .$  For fixed values of the indices the representation space is the fundamental representation of $SU(K/L)$ for finite $K + L$.  Given these variables we can construct matrices 
\begin{equation} M_{ab}^{ij} \equiv q_{ac}^i q_{cb}^j , \end{equation} and take the Hamiltonian to be the trace of a polynomial in these matrices.  The matrices transform like anti-symmetric tensors of any rank and the trace projects out the two forms. Any such Hamiltonian is invariant under $U(N)$ 
transformations of the indices, which are the "fuzzy" analog of area preserving maps.
It's well known that under such a map, a small spherical cap is mapped into an "amoeba" of arbitrary shape with the same area.  Thus, the fact that a region is small just means that it takes time to scramble information injected into a small number of degrees of freedom.  Each action of the Hamiltonian scrambles some finite number of variables, and thus the number of variables reached is exponential in the number of time steps.  There is no suppression of interaction between "distant" regions, because the Hamiltonian doesn't know about the geometry of the sphere but only its measure.  
The time scale of horizon dynamics $ \sim N$ is reproduced If we write
\begin{equation} H_{in} = \frac{1}{N} {\rm Tr}\ P(M/N) , \end{equation} where we have invoked 't Hooft scaling of matrix models.  

We now want to ask about constraints that could correspond to localized jets of particles penetrating the boundaries of the diamond.  Such jets should be non-interacting if $N$ is large, and the obvious (only?) way to do that is to constrain the initial states such that all of the matrices $M$ are block diagonal, with a number of blocks of size $E_k \sim 1$ and one block of size $\sum E_k N$ .   This reproduces (in the sense of scaling with $N$ the dS black hole entropy formula if $\sum E_k$ is identified with a multiple of the asymptotically conserved (in flat space) energy.  Of course in dS space energy is conserved only over times of order $R$.  In fact, the generic Hamiltonian of this form leads to an asymptotically conserved energy if we allow $N \sim t$ as $t \rightarrow\infty$ and reproduces the gross thermal properties of dS space if $N$ is kept large but finite as $t$ is taken to infinity.

We do not have space here to describe the details, but these models define a scattering theory that is manifestly compatible with space-time translation invariance, unitarity and causality.  It's asymptotic Hilbert space has a description as a representation of the Poincare invariant (generalized) AGS algebra.   The constraint of Lorentz invariance, which follows from the HST consistency conditions for geodesics in relative motion, has not yet been implemented.  Experience with string theory leads us to expect that most of the models we've described will not pass this test.

All of the models have meta-stable excitations with all of the scaling properties of entropy, temperature,energy and scrambling properties that we expect\cite{bfbhfw}.  The conditions for black hole formation also meet semi-classical expectations: a black hole is formed when the energy in Planck units inside a causal diamond is of order the Schwarzschild radius in Planck units.  The models all resolve the firewall paradox\cite{fw}: when a system falls onto a black hole the entropy of the relevant Hilbert space increases by a large amount because of the off diagonal matrix elements.  The latter are however initially frozen and the time that it takes to equilibrate the infalling system  is a time over which that system behaves as if it were in flat space.

\section{Holographic Cosmology}

The HST model of cosmology proposes new answers to some of the oldest questions in the subject: how the initial singularity is avoided and why the universe started out in a low entropy state.  There is nothing in the quantum mechanics of time dependent Hamiltonians that insists that any finite point in time must be infinitely far away from past or future time-like infinity.  In HST, merely positing an initial time at some finite time in the past, tells us that if we choose the nested diamonds to all begin at the initial point, then $H_{in} (t_1)$ acts on a Hilbert space of very low dimension.  It cannot resemble the Hamiltonian of a quantum field theory and no features of it exhibit classical hydrodynamic flow.  There's nothing singular about the beginning of the universe, it just isn't described by the classical or quantum Einstein equations.

The Boltzmann-Penrose question has a more interesting resolution.  One can construct a one parameter set of models in which each causal diamond along each geodesic, is always in a maximal entropy state.  The Hamiltonians are simply chosen to be those of the previous section, but we start in a generic state so there are no terms in the Hamiltonian in constrained sectors.  If we allow this evolution to go on forever, but insist that at large time $t$ the Hamiltonian approaches that of a $1 + 1$ dimensional chiral CFT, with central charge $\sim t^2$, and UV cutoff $\sim 1/t$ ,  living on an interval of size $K t$ with $1 \ll K \ll t$.  Since the comoving volume of a flat FRW model scales like $t^3$, the comoving entropy and energy densities scale like \begin{equation} \sigma \sim 1/t   \ \ \ \ \ \ \rho \sim 1/t^2 . \end{equation}  We automatically get the Friedman relation $\rho \sim 1/t^2$ and the characteristic energy/entropy density relation $\sigma \sim \sqrt{\rho}$ of the $p = \rho$ equation of state, which\cite{fsb} showed saturated the early universe entropy bound.  The large number $K$ is related to the range of the indices $i$, while the $t^2$ central charge should be thought of as originating from $t^2$ field $q_{ab}$ in the CFT.

If we modify the model so that after a certain time $N$, the number of $q_{ab}$ does not increase then we get a flat FRW model that interpolates between $p = \rho$ and $p = -\rho$, with no localized excitations at any intermediate time.  These models are completely formulated as consistent quantum theories of gravity.  From the point of view of the asymptotic future dS space of these models, any localized excitation is a constrained state, with the approximately conserved energy $E$ determining the number of constraints.  For fixed $E$, the most probable state to find localized in the bulk is a single black hole of mass $E$.  If we require that the universe have a radiation dominated era, this initial condition is not acceptable.

In HST, the global structure of the universe is not an initial condition.  Homogeneity is a choice of {\it model}: we insist that we have a set of identical quantum systems, which can be associated with different geodesics in the universe.  The Hilbert space of a given system is ${\cal H} (t,{\bf x})$ and its Hamiltonian $H(t, {\bf x})$, and both have the tensor factorization property implied by a nested sequence of causal diamonds corresponding to the nested sequence of intervals $[0,t]$.  An asymptotic dS future implies that as $t\rightarrow\infty$, the dimension of the Hilbert space asymptotes to $e^{c N^2}$.

Now, fix the radius $N$ of the asymptotic dS space and insist that, following the initial, generic, $p = \rho$ beginning, the Hilbert space remains at finite size $1 \ll n \ll N$ for $N_e$ e-folds of proper time.  That is, we have an inflationary era, with inflationary Hubble parameter $n^{-1}$ in Planck units.  We then let the horizon expand gradually, by changing the size of the Hilbert space on which $H_{in} (t)$ acts.  This is what corresponds to the slow roll era. As the horizon expands, we must satisfy the QGCI constraints.  Geodesics originally outside the inflationary horizon volume enter into it (Figure 3).  This occurs on FRW time slices prior to the end of slow roll.  Let's label one particular geodesic as the origin , ${\bf 0}$. During the period up to horizon entry, $H_{out} (t, {\bf 0})$ along one geodesic must coincide with $H_{in} (t, {\bf x})$ for the trajectory just entering the horizon, for all $t$ prior to horizon entry.  Since $t$ is in the inflationary regime, the latter operator acts on a Hilbert space of entropy $n^2$ and the state is random. This means that the initial conditions in the out part of the Hilbert space ${\cal H} (t, {\bf 0})$ must be such that there is a tensor factor of entropy $n^2$ which has not interacted with the rest of the system.  Given the form of the Hamiltonian this implies that the initial conditions are constrained, and there is a localized object of entropy $n^2$ coming into the horizon.  Since the {\it model} is defined by a homogeneous and isotropic collection of quantum systems, we get a homogeneous and isotropic gas of black holes on FRW slices of the space-time.  Homogeneity and isotropy cannot be exact however, because the HST consistency conditions only constrain the density matrix in the intersection of two diamonds.  There are thus statistical fluctuations, in the entropy of the black holes and in their angular momentum.  These are, by the usual rules of statistical mechanics, primarily Gaussian, with a size $\delta H/H \sim 1/n$, the inverse square root of the entropy.  Classical cosmological perturbation theory gives the conventional gauge invariant measure of fluctuations as $\zeta = \delta H/H\epsilon $, where $\epsilon = \dot{H}{H^2} $ is the slow roll factor.  Non-gaussianity is down by inverse powers of $n$.  

\begin{figure}[h!]
\begin{center}
\includegraphics[width=12cm]{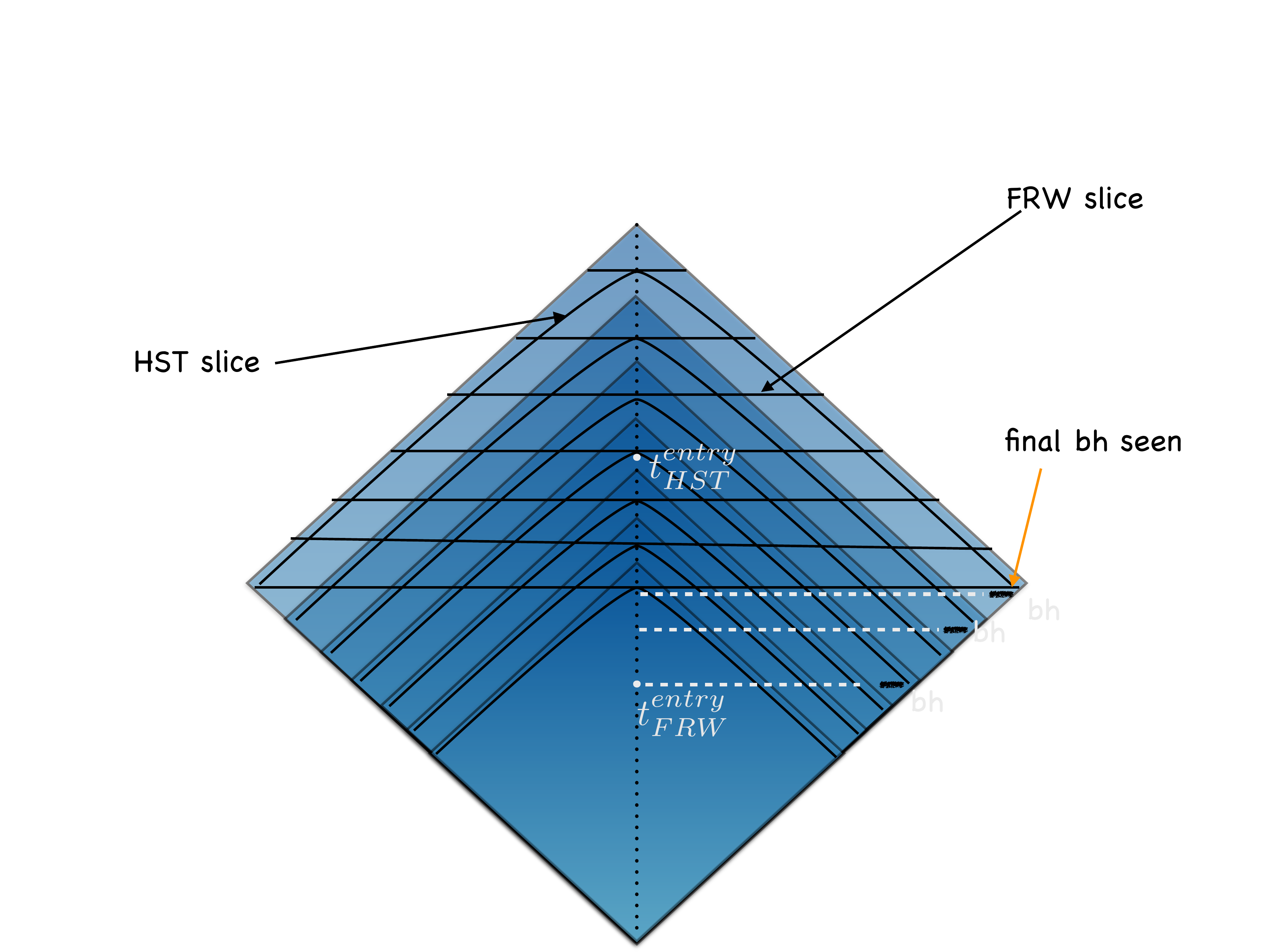}
\end{center}

\caption{ 
black holes entering horizon}

\label{fig3}
\end{figure}

The detailed form\cite{holocosmrevisedGRessay} of the predictions of this model is quite different than that of conventional inflation, particularly for tensor fluctuations and non-Gaussianities, but for scalar fluctuations the predictions depend on the slow roll factor $\epsilon (t)$.  The only observational evidence we have constraining $\epsilon (t)$ comes from scalar CMB fluctuations, so it's obvious that this model can fit the existing data.  Nominally, the ratio of tensor to scalar fluctuations is of order $r \sim \epsilon^2$, and this suggests it is very small, but we have not yet been able to calculate the "$o(1)$" coefficients in this ratio.   So far we have found no theoretical constraint on the form of $\epsilon (t)$ except for the fact that, as is clear from Figure 3, inflation lasts half the conformal time of the universe (which is bounded because we have an asymptotic dS space).  This gives a total number of e-folds as $\sim 80$.   Approximate $SO(1,4)$ invariance of the fluctuations is obtained by combining the approximate $SL(2,R)$ of individual horizon volumes with the $SO(3)$ of any given geodesic.

\section{Conclusions}

Bekenstein's revolutionary association of entropy with black hole area provides the kernel, in our opinion, for the entire quantum theory of gravity.  Quantum field theory calculations by Hawking and others, which validate Bekenstein's idea, do not provide us with a model for the microscopic degrees of freedom responsible for that entropy.
The success of those calculations is based on the geometric fact that the analytic continuation of a horizon to imaginary time contains a one cycle of fixed length at infinity. Unruh's result on accelerated trajectories in flat space-time uses the same geometric observation and suggests that Bekenstein's idea applies to more general regions in space-time, and that states that quantum field theory treats as unique vacuum states, actually have large entropy.  The work of Jacobson, Fischler-Susskind, and Bousso, gives us a natural extension of Bekenstein's conjecture to arbitrary causal diamonds in arbitrary space-times. Jacobson's work in particular shows that most of the states associated with the BHJFSB entropy, have very small energy when viewed from the geodesic coordinate system in a given causal diamond.  The Schwarzschild de Sitter entropy formula then shows us that states localized in a causal diamond in de Sitter or asymptotically flat space, are constrained states of the fundamental degrees of freedom, with the conventional energy equal to the number of constraints divided by the dS radius.  

The HST formalism is a general proposal for quantum theories of gravity in arbitrary space-times, based on the ideas outlined in the previous paragraph.  A classical space-time gives rise to a set of causal diamonds.  HST is based on combining the field theoretic idea that any such diamond corresponds to a Murray-von Neumann factor in the complete operator algebra of space-time.  The BHJFSB entropy formula implies that most Hilbert spaces involved in quantum gravity are finite dimensional.  The exceptions among "commonly encountered" space-times are the asymptotic Hilbert spaces of asymptotically flat and AdS space-times.  These are two, quite different limits of the physics in finite causal diamonds.  We have argued elsewhere\cite{tbwfads} that the CFT boundary Hamiltonian of AdS space-times does not describe time evolution in causal diamonds much smaller than the AdS radius, except for a small set of states, in perturbation theory. Minkowski space physics is a straightforward extension of the physics of localized objects in finite causal diamonds and is approximately the same as the de Sitter physics of such objects over time scales short compared to the dS radius\footnote{For bound clusters of objects, which have "decoupled from the Hubble flow" until they collapse into black holes and decay, the approximation remains valid for a much longer time.  Deviations from it come only from thermal "accidents" in dS space, and from jets of radiation emitted from the cluster, which enter the horizon after a de Sitter time.}

Much of the work on HST, including the cosmological models of the penultimate section, does not depend heavily on the precise form of the variables or Hamiltonians that we use to describe the quantum systems underlying classical space-times.  This is because of Bekenstein's lesson that the situations in quantum gravity that are poorly described by quantum field theory, have a coarse grained thermodynamic description, which is adequate to predict most of what we need to account for current observations.  It would not be amiss to designate the early universe in the HST model as a {\it Bekensteinian Cosmology}, dominated by hydrodynamics and the statistical fluctuations of systems with large but finite entropy.

\vfill\eject
\vskip.3in
\begin{center}
{\bf Acknowledgments }\\
The work of T.Banks is {\bf\it NOT} supported by the Department of Energy, the NSF, the Simons Foundation, the Templeton Foundation or FQXI. The research of WF is based upon work supported by the National Science Foundation under Grant Number PHY-1620610.\end{center}

\end{document}